\documentclass[aps,twocolumn,pra,
preprintnumbers,amsmath,amssymb,superscriptaddress]{revtex4}
\usepackage{graphicx}
\usepackage{color}
\usepackage{url}

\definecolor{refkey}{rgb}{0.9451,0.2706,0.4941}
\definecolor{labelkey}{rgb}{0.9451,0.2706,0.4941}

\newcommand{\eq} {equation}
\newcommand{\eqa} {eqnarray}
\newcommand{\NN} {\nonumber}

\newcommand{\tr} {\mathrm{tr}}
\newcommand{\wt} {\widetilde}

\usepackage{amsmath}	
\begin{document}


\title{
Resumming perturbative series
in the presence of monopole bubbling effects
}

\date{\today}

\author{Masazumi Honda}\email[]{masazumi.hondaATweizmann.ac.il} 
\affiliation{{\it Department of Particle Physics, Weizmann Institute of Science, Rehovot 7610001, Israel}}

\author{Daisuke Yokoyama}\email[]{dyokoyamaATfudan.edu.cn} 
\affiliation{{\it Department of Mathematics, Kings College London,
The Strand, London, WC2R 2LS, U.K.\\
and
Department of Physics and Center for Field Theory and Particle Physics, Fudan University, 220
Handan Road, 200433 Shanghai, China}}

\begin{abstract}
Monopole bubbling effect
is screening of magnetic charges of singular Dirac monopoles by regular 't Hooft-Polyakov monopoles.
We study properties of weak coupling perturbative series
in the presence of monopole bubbling effects as well as instantons.
For this purpose,
we analyze supersymmetric 't Hooft loop 
in four dimensional $\mathcal{N}=2$ supersymmetric gauge theories
with Lagrangians and non-positive beta functions.
We show that 
the perturbative series of the 't Hooft loop 
is Borel summable along positive real axis for fixed instanton numbers and screened magnetic charges.
It turns out that
the exact result of the 't Hooft loop is the same as
the sum of the Borel resummations 
over instanton numbers and effective magnetic charges.
We also obtain the same
result for supersymmetric dyonic loops.
\end{abstract}

\maketitle



\noindent

\section{Introduction}
Progress on quantum field theory (QFT) hinges on understanding non-perturbative effects
such as instantons, monopoles, vortices and so on.
One of much less understood non-perturbative effects is monopole bubbling effect,
which is screening of magnetic charges of singular Dirac monopoles \cite{Kapustin:2006pk}
by regular 't Hooft-Polyakov monopoles \cite{tHooft:1974kcl} 
\footnote{See \cite{Cherkis:2007jm} for an explicit example.}.
Aim of this paper is
to understand properties of weak coupling perturbative series
in the presence of monopole bubbling effects as well as instantons.

The monopole bubbling effects
typically appear in 't Hooft loop \cite{tHooft:1977nqb},
which is magnetic version of Wilson loop
and detects Higgs phase by area law,
while area law of Wilson loop implies confinement \cite{Wilson:1974sk}.
The 't Hooft loop can be defined as a partition function 
with singular boundary conditions for gauge fields
given by Dirac monopoles \cite{Kapustin:2005py}.
For example,
a straight 't Hooft line along the Euclidean time 
at spatial origin in $\mathbb{R}^4$ is described by
\begin{\eq}
F = \frac{B}{4} \epsilon_{\mu\nu\rho} \frac{x^\mu}{|x|^3} dx^\nu \wedge dx^\rho ,
\end{\eq}
where $B$ is a flux on $S^2$ surrounding the loop $B=\frac{1}{2\pi} \int_{S^2} F$.
It is known that 
't Hooft loop receives not only instanton corrections but also monopole bubbling effects.
In this paper 
we study perturbative series of half-BPS 't Hooft loops 
in 4d $\mathcal{N}=2$ supersymmetric (SUSY) gauge theories.
To preserve SUSY, 
we need an additional boundary condition for 
the adjoint scalar $\Phi$  in the $\mathcal{N}=2$ vector multiplet: $\Phi = \frac{B}{2|x|}$.
Here we analyze the SUSY 't Hooft loops put on a curved space.

Before going to our setup in detail,
let us recall general expectations on perturbative series in QFT.
Perturbative series in QFT is typically not convergent \cite{Dyson:1952tj}.
In mathematics,
there is a standard way to resum non-convergent series called Borel resummation.
Given a perturbative series 
$\sum_{\ell =0}^\infty c_\ell g^{a+\ell} $
of a quantity $I(g)$,
its Borel resummation along $\mathbb{R}_+$ is defined by
\begin{\eq}
\mathcal{S}_0 I (g)
=\int_0^{\infty } dt\ e^{-\frac{t}{g}} \mathcal{B}I(t) ,
\label{eq:Borel} 
\end{\eq}
where $\mathcal{B}I(t)$ is an analytic continuation of 
the formal Borel transformation  
$\sum_{\ell =0}^\infty 
\frac{c_\ell}{\Gamma (a+\ell )} t^{a+\ell -1}$
after the summation.
Perturbative series in typical QFT is expected to be
non-Borel summable due to singularities along $\mathbb{R}_+$ of $\mathcal{B}I(t)$ 
and Borel resummation formula has ambiguities
since the integral depends on how to avoid the singularities.
However, 
it is generally unclear when perturbative series in QFT are Borel summable.
Another important question is 
in the case that we do not have Borel ambiguities,
how the resummation is related to exact results.

In \cite{Honda:2016mvg}
one of the authors initiated to address these questions.
It has been proven that
perturbative series 
in 4d $\mathcal{N}=2$ and 5d $\mathcal{N}=1$ SUSY gauge theories 
with Lagrangians
are Borel summable along $\mathbb{R}_+$ for various observables in sectors with fixed instanton numbers
\footnote{
See \cite{Russo:2012kj,Aniceto:2014hoa} for earlier checks in some examples.    
}.
This result for the 4d $\mathcal{N}=2$ case
is expected from a recent proposal \cite{Argyres:2012vv}
on a semi-classical realization of IR renormalons \cite{tHooft:1977xjm},
where the semi-classical solution does not exist 
in the 4d $\mathcal{N}=2$ theories \cite{Poppitz:2011wy} (see also \cite{Dunne:2015eoa}).
In this paper we study the 't Hooft loop in the 4d $\mathcal{N}=2$ theories,
which receives monopole bubbling effects in addition to instanton effects.

Now we explain details of our setup.
Let us consider 4d $\mathcal{N}=2$ theories
with Lagrangians and non-positive beta functions.
In this class of theories
we study the supersymmetric 't Hooft loop
placed at a circle in a squashed sphere $S^4_b$,
which is defined as
the hypersurface in $\mathbb{R}^5$
\begin{\eq}
X_0^2 +b^{-2}(X_1^2 +X_2^2 )  +b^2 (X_3^2 +X_4^2 )  = r^2 .
\end{\eq}
In this geometry,
we can place half-BPS line operators
on the circles $S^1_b$ and $S^1_{b^{-1}}$,
which are given by $X_3 =X_4 =0$ and $X_1 =X_2=0$ at fixed $X_0$, respectively.
Though we focus on $S^1_b$ at $X_0 =0$,
the result for $S^1_{b^{-1}}$
is simply obtained by the replacement $b\rightarrow b^{-1}$,
which does not change our main result.

The 't Hooft loop receives 
both instantons and monopole bubbling effects \cite{Gomis:2009ir} 
\footnote{
Note that because of $|v|\leq |B|$,
the contributions from the sector $|v|\neq |B|$ 
are relatively exponentially suppressed corrections.
}:
\begin{\eq}
T(B)
= \sum_{k,\bar{k},v} e^{-kS_{\rm inst}-\bar{k}\bar{S}_{\rm inst} 
-{\rm tr}(v^2 ) S_{\rm mono}} T^{(k,\bar{k},v)} (g,\theta ) ,
\label{eq:expansion}
\end{\eq}
where $v$ denotes a screened monopole charge and
\begin{\eq}
S_{\rm inst}=-2\pi i\tau ,\quad
\bar{S}_{\rm inst}=2\pi i\bar{\tau},\quad
S_{\rm mono}= -\frac{\pi b^2 }{2g} ,
\end{\eq}
with a complex gauge coupling $\tau =\theta /2\pi +i/g$.
Eq.~\eqref{eq:expansion} is schematic
in the sense that theories with product gauge group
have multiple couplings and multiple $(k,\bar{k},v)$.
Here we are interested in weak coupling expansion by (square) Yang-Mills coupling $g$:
\begin{\eq}
T^{(k,\bar{k},v)} (g,\theta )
\simeq \sum_\ell c^{(k,\bar{k},v)}_\ell (\theta ) g^{\sharp +\ell} ,
\end{\eq}
where $\sharp$ is a leading order exponent.
In this paper 
we show that
the perturbative series in the sector with fixed $(k,\bar{k},v )$ is Borel summable along $\mathbb{R}_+$
\footnote{
\cite{Russo:2012kj} studied the 't Hooft loop in $SU(2)$ $\mathcal{N}=2^\ast$ Yang-Mills theory
on round $S^4$ in perturbative sector.
}
in the case that
explicit expressions for instantons and monopole bubbling effects 
in SUSY localization formula \cite{Pestun:2007rz}
are available.
Our main result is
\begin{\eq}
 T^{(k,\bar{k},v)} (g,\theta )
=\mathcal{S}_0  T^{(k,\bar{k},v)} (g,\theta ) .
\end{\eq}
Thus we can rewrite the whole exact result 
in terms of the Borel resummation:
\begin{\eq}
T(B)
= \sum_{k,\bar{k},v} e^{-kS_{\rm inst}-\bar{k}\bar{S}_{\rm inst} 
-{\rm tr}(v^2 ) S_{\rm mono}} 
\mathcal{S}_0  T^{(k,\bar{k},v)} (g,\theta ) .
\end{\eq}
This equation can be understood 
as ``semi-classical decoding" \cite{decoding}
of the 't Hooft loop in 4d $\mathcal{N}=2$ theories
though the resurgence structure itself is trivial for this case.

\section{'t Hooft loop and Borel resummation}
\label{sec:proof}
\subsubsection*{Exact result}
Instead of a path integral expression,
we use a conjectural finite dimensional integral representation 
for the 't Hooft loop \cite{Ito:2011ea,Gomis:2011pf,Okuda:2014fja}
which has passed highly nontrivial tests.
There are exact results for the 't Hooft loop
in the 4d $\mathcal{N}=2$ theories on round $S^4$ \cite{Gomis:2011pf}, 
$S^1 \times\mathbb{R}^3$ \cite{Ito:2011ea}
and $S^1 \times S^3$ \cite{Gang:2012yr}
by SUSY localization \cite{Pestun:2007rz}.
Although there is no explicit computation for our $S_b^4$ case, 
one can find the reasonable expression for the exact result
by combining the results for round $S^4$ \cite{Gomis:2011pf}, 
$S^1 \times\mathbb{R}^3$ \cite{Ito:2011ea}
and the partition function on $S_b^4$ \cite{Hama:2012bg}
(see also \cite{Nosaka:2013cpa}).
The expression is consistent 
with the exact results on round $S^4$ for $b=1$, the partition function on $S_b^4$ for $B=0$,
and AGT relation \cite{Alday:2009aq} for some theories.

Let us consider the 4d $\mathcal{N}=2$ SUSY gauge theories
with a semi-simple gauge group $G=G_1 \times\cdots\times G_n$,
which are coupled to
hyper multiplets of representations $(\mathbf{R_1} ,\cdots ,\mathbf{R_{N_f}} )$.
According to \cite{Ito:2011ea,Gomis:2011pf,Okuda:2014fja}, 
the 't Hooft loop is given by 
\footnote{We take $r=1$. 
We can recover $r$-dependence by taking $a\rightarrow ra$.}
\begin{\eq}
T(B)
=\sum_v \int d^{|G|}a 
Z_{\rm cl}^{(v)} Z_{\rm 1loop}^{(v)} Z_{\rm NP}^{(v)} ,
\label{eq:MM}
\end{\eq}
where $v$ describes the screened magnetic charges.
The classical contribution $Z_{\rm cl}^{(v)} (a)$ is
\begin{\eq}
Z_{\rm cl}^{(v)} (a)
=\prod_{p=1}^n \exp\Biggl[ 
- \frac{{\rm tr}_p a^2}{g_p}  -b \theta_p {\rm tr}_p (a\cdot v ) 
\Biggr] ,
\end{\eq}
where $g$ is proportional to square of one-loop effective Yang-Mills coupling at scale $1/r$
and ${\rm tr}_p$ denotes trace in the gauge group $G_p$.
The one-loop determinant $Z_{\rm 1loop}^{(v)}$ has contributions 
both from poles and equator of $S_b^4$ 
\footnote{
Note that $\alpha\cdot v$ and $\rho\cdot v$ are integers
by Dirac quantization conditions \cite{Goddard:1976qe}.
}:
\begin{\eqa}
&& Z_{\rm 1loop}^{(v)} (a)
= \left| Z_N (a_N ) \right|^2 Z_{\rm eq}(a ) , \NN\\
&& Z_N (a)
=\Biggl[
 \frac{\prod_{\alpha \in \Delta_+} \Upsilon (i \alpha\cdot a ) \Upsilon (-i \alpha\cdot a ) 
 }
{\prod_{I=1}^{N_f} \prod_{\rho_I \in \mathbf{R_m}}
 \Upsilon \left( i \rho_I \cdot a +\frac{Q}{2} \right)  } \Biggr]^{1/2} , \NN\\
&& Z_{\rm eq} (a) \NN\\
&& =
\frac{\prod_{I,\rho_I}  \prod_{k=0}^{|\rho_I \cdot v|-1}
\cosh^{\frac{1}{2}}{\Bigl[ \pi b \left(\rho_I \cdot a 
+i b \left( k-\frac{|\rho_I \cdot v|-1}{2}\right) \right) \Bigr]} }
{\prod_{\alpha \in \Delta} \prod_{k=0}^{|\alpha \cdot v|-1}
\sinh^{\frac{1}{2}}{\Bigl[ \pi b\left(\alpha \cdot a 
+ib \left( k-\frac{|\alpha \cdot v|}{2} \right) \right)\Bigr]}  }
,\NN\\
\end{\eqa}
where $Q=b+b^{-1}$, $a_N =a +\frac{ibv}{2}$, and
$\Upsilon (x)$ is the Upsilon function defined by
\begin{\eq}
\Upsilon (x) 
=\prod_{m_1 ,m_2 \geq 0} (m_1 b +m_2 b^{-1} +x)(m_1 b +m_2 b^{-1} +Q-x) .
\label{eq:Upsilon}
\end{\eq}
$Z_{\rm NP}^{(v)} $ expresses non-perturbative effects, which are more involved:
\begin{\eqa}
&& Z_{\rm NP}^{(v)}(a)
=Z_{\rm inst}^{(v)} (a) Z_{\rm mono}^{(v)} (a) ,\NN\\
&& Z_{\rm inst}^{(v)}
=|Z_{\rm Nek}(a_N )|^2 
=\sum_{k,\bar{k}} e^{-kS_{\rm inst}-\bar{k}\bar{S}_{\rm inst} } 
Z_{\rm inst}^{(k,\bar{k},v)} ,
\end{\eqa}
where $Z_{\rm Nek}^{(v)}$ is Nekrasov instanton partition function \cite{Nekrasov:2002qd}
with $\Omega$-background parameters $(\epsilon_1 ,\epsilon_2 )=(b,b^{-1})$,
and $Z_{\rm mono}^{(v)}$ is a contribution from monopole bubbling effects.
Note that $Z_{\rm mono}^{(v)}(a)$ for $v=B$ is trivial:
\begin{\eq}
 Z_{\rm mono}^{(B)}(a) =1 ,
 \label{eq:trivialMono}
\end{\eq}
which reflects the absence of the monopole screening.
Now we are interested 
in properties of small-$g$ expansion of
\begin{\eq}
 T^{(k,\bar{k},v)} (g,\theta )
= \int_{-\infty}^\infty d^{|G|}a \ Z_{\rm cl}^{(v)}Z_{\rm 1loop}^{(v)} Z_{\rm NP}^{(k,\bar{k},v)} ,
\end{\eq}
where
\begin{\eq}
Z_{\rm NP}^{(k,\bar{k},v)} (a)
=Z_{\rm inst}^{(k,\bar{k},v)}(a) Z_{\rm mono}^{(v)} (a) .
\end{\eq}

\subsubsection*{$SU(N)$ superconformal QCD}
We first discuss 4d $\mathcal{N}=2$ $SU(N)$ superconformal QCD
for simplicity of explanations
which is SUSY QCD (SQCD) with $2N$ fundamental hyper multiplets.
We will consider more general theories later.
The 't Hooft loop of our SQCD in the sector $(k,\bar{k},v)$ is
\begin{\eqa}
&& T_{\rm SQCD}^{(k,\bar{k},v)} (g,\theta ) 
 =  \int d^N a\  \delta \Bigl( \sum_j a_j \Bigr)
e^{-\frac{1}{g}\sum_j a_j^2 -b\theta \sum_j v_j a_j} \NN\\
&& 
\prod_{1\leq i< j \leq N}
\frac{\left| \Upsilon (ia_{N,ij} )  \Upsilon (-ia_{N,ij} ) \right|}
{ \prod_{k=0}^{|v_{ij}|-1}
\left| \sinh{\Bigl[  \pi b \left( a_{ij} 
+ib \left( k-\frac{|v_{ij}|}{2} \right) \right)\Bigr]} \right|  }  \NN\\
&& \prod_{j =1}^N \frac{ \prod_{k=0}^{|v_j |-1}
\cosh^{N}{\Bigl[ \pi b \left( a_j 
+ib \left( k-\frac{|v_j|-1}{2}\right) \right) \Bigr]} }
{ \left| \Upsilon \left( i a_{N,j} +\frac{Q}{2} \right) \right|^{2N}
  } 
Z_{\rm NP}^{(k,\bar{k},v)} , \NN\\
\label{eq:exactSQCD}
\end{\eqa}
where $a_{ij}=a_i -a_j$ and $v_{ij}=v_i -v_j$.
Let us study small-$g$ expansion of this object.
Instead of explicitly computing perturbative coefficients,
we explicitly find Borel transformation, somehow already hidden in eq.~\eqref{eq:exactSQCD}
as in \cite{Honda:2016mvg,Honda:2016vmv}.
To see this,
first we make a change of the variables as $a_j =\sqrt{t}\hat{x}_j$,
where $\hat{\mathbf{x}}=(\hat{x}_1 ,\cdots ,\hat{x}_N)$ 
is the unit vector in $\mathbb{R}^N$.
Then, we rewrite the 't Hooft loop as
\begin{\eq}
T_{\rm SQCD}^{(k,\bar{k},v)} (g,\theta )
= \int_0^\infty dt\ e^{-\frac{t}{g}} f^{(k,\bar{k},v)} (t,\theta ) ,
\label{eq:Laplace}
\end{\eq}
where
\begin{\eqa}
&& f^{(k,\bar{k},v)} (t )
=t^{\sharp}\int_{S^{N-1}}d^{N-1}\hat{x} \delta \Bigl( \sum_j \hat{x}_j \Bigr)
h^{(k,\bar{k},v)} (t,\hat{x} ) ,\NN\\
&&  h^{(k,\bar{k},v)} (t,\hat{x} ) 
=\left. e^{-b\theta \sum_j v_j a_j} 
 Z_{\rm 1loop}^{(v)}  Z_{\rm NP}^{(k,\bar{k},v)} \right|_{a_j =\sqrt{t}\hat{x}_j} .
\end{\eqa}
The exponent $\sharp$ is a constant depending on $N$ and $v_{ij}$, 
whose detail is not important for our purpose
\footnote{
Explicitly, $\sharp =\frac{N-2}{2} 
+\sum_{i<j} \delta_{v_{ij} ,0}$.
}.
The functions $f^{(k,\bar{k},v)}$ and $h^{(k,\bar{k},v)}$ depend on $\theta$ 
but we do not often write $\theta$ explicitly in the arguments for simplicity.
Since the expression \eqref{eq:Laplace} is the form of the Laplace transformation
and similar to the Borel resummation,
one may wonder whether the function $f^{(k,\bar{k},v)} (t)$
of the perturbative series is the Borel transformation.
Indeed, we can prove that
this is the case as in \cite{Honda:2016mvg,Honda:2016vmv}:
\begin{\eq}
f^{(k,\bar{k},v)} (t,\theta )
=\mathcal{B}T_{\rm SQCD}^{(k,\bar{k},v)} (t,\theta ) .
\label{eq:SQCDBorel}
\end{\eq}
The proof takes the following three steps.
(1) 
We show that
the integrand $h^{(k,\bar{k},v)}(t,\hat{x})$ is 
identical to analytic continuation of a convergent power series of $t$.
(2) 
We exchange the order of
the power series expansion of $h^{(k,\bar{k},v)}(t,\hat{x})$ 
and integration over $\hat{x}$.
This is rigorously justified 
by proving uniform convergence of the small-$t$ expansion.
(3) 
It is easily seen that
the coefficient of the perturbative series of $f^{(k,\bar{k},v)}(t)$ 
is given by $c_\ell^{(k)} /\Gamma (\frac{\sharp +\ell}{2})$
as guaranteed by the Laplace transformation \eqref{eq:Laplace}. 

First, let us focus on perturbative sector $(k,\bar{k},v)=(0,0,B)$.
We will consider non-perturbative effects later.
To prove the uniform convergence of the small-$t$ expansion of $h^{(k,\bar{k},v)}(t,\hat{x})$,
it is convenient to use Weierstrass M-test,
i.e.
we find a sequence $\{ M_\ell \}$ such that
$| h^{(0)}_\ell (\hat{x})| < M_\ell $ and $\sum_{\ell =0}^\infty M_\ell t^{\sharp +\ell} <\infty $ for fixed $t$.
We can easily find such a series
as in \cite{Honda:2016mvg,Honda:2016vmv}
by replacing all the sources of negative contributions
to the small-$t$ expansion by positive definite larger values
\footnote{
For example, 
it is convenient to make the replacements 
$\pm \hat{x}_i \rightarrow 1$, $\pm\hat{x}_{ij}\rightarrow 2$, 
$\zeta (\ell >1 )\rightarrow 2$ and
$1/[(m_1 b +m_2 b^{-1})^2 +x^2 ]$ $\rightarrow$
$1/[b^{\pm 2}(m_1  +m_2 )^2 +x^2 ]$.
}.
As an example,
the following function generates $M_\ell$:
\begin{\eqa}
\frac{P(t) e^{b\theta \sqrt{t}\sum_j |v_j | -4(5N^2 -4N) b^{2}t }}{(1 - b^{2}t )^{4N^2}
\prod_\pm \left( 1-2b^{2}(2t \pm \sqrt{t}Q) \right)^{2(N^2 -N)} } ,
\end{\eqa}
where $P(t)$ is an appropriate finite order polynomial.
Thus, $f^{(0,0,B)} (t )$ is actually the Borel transformation.

Now we can explicitly study 
analytic properties of the Borel transformation.
Since we have expressed
the Borel transformation in terms of the one-loop determinant $Z_{\rm 1loop}^{(v)}$,
this problem boils down 
to analytic properties of the one-loop determinant,
whose details are studied in app.~\ref{app:analytic} for general case.  
As a result, 
the one-loop determinant in our SQCD has degree-$2N$ poles at
\begin{\eq}
a_j
= \pm i \left[  \left( m_1 +\frac{|v_j |+1}{2} \right) b +\left(m_2 +\frac{1}{2} \right) b^{-1} \right]  ,
\end{\eq}
where $m_1 ,m_2 \in \mathbb{Z}_{\geq 0}$.
We emphasize that
all the apparent branch cuts are canceled in the whole expression.
Most important point here is that
the one-loop determinant does not have singularities for $a\in\mathbb{R}$.
Thus, the Borel transformation 
has singularities (poles) only along $\mathbb{R}_-$ for $b\in\mathbb{R}$
and the perturbative series is Borel summable along $\mathbb{R}_+$.

\subsubsection*{General $\mathcal{N}=2$ theories with Lagrangian}
We can easily generalize the above analysis
to general $\mathcal{N}=2$ theories with Lagrangian.
Taking the polar coordinate $a_i^{(p)}=\sqrt{t_p} \hat{x}_i^{(p)} $ 
with $\hat{x}_i^{(p)}\in S^{|G_p |-1}$,
we find 
\footnote{If $G_p$ is $SU$, then we insert the traceless constraint by the delta function into $h^{(k,\bar{k},v)}$.}
\begin{\eq}
T^{(k ,\bar{k}, v )} (g,\theta ) 
= \int_0^\infty d^n t\  e^{-\sum_{p=1}^n \frac{t_p}{g_p}}
 f^{(k ,\bar{k}, v )}  (t) ,
\label{eq:multiple}
\end{\eq}
where 
\footnote{
$(k,\bar{k},v)$ stands for $k=\{ k_1 ,\cdots ,k_p \}$, $\bar{k}=\{ \bar{k}_1 ,\cdots ,\bar{k}_p \}$
and $v=\{ v^{(1)}, \cdots ,v^{(p)} \}$.
}
\begin{\eqa}
&& f^{(k,\bar{k},v )} (t )
= t^{\sharp} \int_{\rm sphere}d\hat{x}\  
h^{( k,\bar{k},v)} (t,\hat{x} ) ,\NN\\
&& \left. h^{(k,\bar{k},v )} (t,\hat{x} ) 
= e^{-b\sum_p \theta_p {\rm tr}_p v\cdot a} 
 Z_{\rm 1loop}^{(v )}  Z_{\rm NP}^{(k,\bar{k},v )}
  \right|_{a_i^{(p)}=\sqrt{t_p} \hat{x}_i^{^{(p)}} } ,  \NN\\
\label{eq:f_multi} 
\end{\eqa}
where $t^\sharp =\prod_{p=1}^n t_p^{\sharp_p} $
with a constant $\sharp_p$ unimportant for our purpose
\footnote{
Explicitly, $\sharp_p =\frac{|G_p|-1}{2} 
+\sum_{\alpha_p \in \Delta_+^{(p)}} \delta_{\alpha_p \cdot v^{(p)} ,0}$.
}.
Again, eq.~\eqref{eq:multiple}
is similar to the form of Borel resummation generalized to multiple couplings.
Indeed, one can show that
the function $f^{( k,\bar{k},v )} (t )$ is 
nothing but the Borel transformation of the perturbative series.
Let us focus on the sector with $(k,\bar{k},v)=(0,0,B)$ again.
As in the previous case,
we can always show that
$f^{( k,\bar{k},v )} (t )$ is the Borel transformation:
\begin{\eq}
\mathcal{B}T^{(\{k ,\bar{k}, v \})} (t,\theta )  
= f^{(\{k ,\bar{k}, v \})}  (t,\theta ) .
\label{eq:fBorel}
\end{\eq}

Let us look at analytic properties of the Borel transformation.
According to app.~\ref{app:analytic},
the one-loop determinant has singularities at
\begin{\eq}
\rho\cdot a
= \pm i \left[  \left( m_1 +\frac{|\rho\cdot v|+1}{2} \right) b +\left(m_2 +\frac{1}{2} \right) b^{-1} \right]  .
\end{\eq}
This shows that
the one-loop determinant does not have singularities for real $a$
and therefore the perturbative series is Borel summable along positive real axis.

\subsubsection*{Instanton corrections}
Generalization to non-zero instanton sector is also straightforward
because
$Z_{\rm Nek}(a )$ with fixed instanton number 
and omega background parameters $(\epsilon_1 ,\epsilon_2 )=(b,b^{-1})$ is
a rational function, which does not have poles for real $a$
unless $m_1 b +m_2 b^{-1}$ ($m_{1,2} \in\mathbb{Z}$)
can be purely imaginary.
Therefore, eq.~\eqref{eq:fBorel} still holds in nonzero instanton sector.

For example, 
Nekrasov partition function for $U(N)$ SQCD with $N_f$ fundamental hypers is given by
\begin{\eq}
Z_{\rm Nek}(a)
=\sum_Y  e^{-|Y| S_{\rm inst}}
\frac{\left( \prod_{j=1}^N  n^f_j (a,Y)  \right)^{N_f} }
     {\prod_{i,j =1}^N n^V_{i,j} (a ,Y ) } ,
\end{\eq}
where $Y=(Y_1 ,\cdots ,Y_N )$ denotes a set of Young tableau in $U(N)$,
$|Y|$ is the total number of boxes of $Y$,
and
\begin{\eqa}
&& n^V_{i,j}(a, Y ) 
= \prod_{ s\in Y_i} E_{ij}(a ,s) ( Q -E_{ij}(a ,s) ) ,\NN\\
&& E_{ij} (a ,s)
= -bA_{Y_j} (s )  +b^{-1}( L_{Y_i} (s)  +1 )  -ia_{ij}  ,\NN\\
&& n^f_j (a,Y ) 
= \prod_{s\in Y_j } \phi_j ( a,s ) (\phi_j (a,s)  +Q )   ,\NN\\
&& \phi_j (a,s) 
= -ia_j  +b ( s_h -1)   +b^{-1}( s_v -1 ) .
\end{\eqa}
The indices $s=(s_h ,s_v )$ label
a box in Young diagram at $s_h$-th column and $s_v$-th row,
and $L_Y (s)$ ($A_Y (s)$) is leg (arm) length 
of Young tableau $Y$ at $s$.
The vector contribution gives poles at
\begin{\eq}
a_{ij}
= -ib \left( \frac{v_{ij}}{2} -A_{Y_j} (s)  \right) 
  -ib^{-1} \left( L_{Y_i}(s)+1\right) ,
\end{\eq}
and
\begin{\eq}
a_{ij}
= ib \left( 1-\frac{v_{ij}}{2} +A_{Y_j} (s)  \right) 
  -ib^{-1}L_{Y_i}(s) .
\end{\eq}
Although the contribution from each Young diagram may have poles at $a_{ij} =0$,
this type of poles are canceled 
after summing over Young diagrams 
with the same instanton number \cite{Pestun:2007rz}.
Thus the Borel transformation has singularities
only along $\mathbb{R}_-$
and especially perturbative series is Borel summable along $\mathbb{R}_+$
even if we include the instanton corrections.

\subsubsection*{Monopole bubbling effects}
Now let us add monopole bubbling effects.
As in the instanton corrections,
generalization is straightforward 
as long as we know explicit expressions of $Z_{\rm mono}^{(v)}(a)$
because the monopole bubbling effect is described by a ratio of finite products of hyperbolic functions,
whose poles are not on real axis.
Hence, \eqref{eq:fBorel} still holds in the presence of the monopole bubbling effects.

For example,
for $SU(N)$ or $U(N)$ gauge theory with fundamentals and adjoints,
the contribution from the monopole bubbling effects is given by
\begin{\eq}
Z_{\rm mono}^{(v)}(a)
 = \hat{\sum_{Y}}   Z_Y^{\rm vec} (a) \prod_I Z_Y^{R_I} (a) ,
 \label{eq:mono}
\end{\eq}
where
\begin{\eqa}
&& Z_Y^{\rm vec} (a)
= \frac{1}
{\hat{\prod}_{i, j, s\in Y_i, \pm} 
\sinh{\frac{2\pi ba_{ij} +i\pi b^2(A_{Y_i}(s) -L_{Y_j}(s) \pm 1 )}{2}}  }  ,\NN\\
&& Z_Y^{\rm adj} (a)
= \hat{\prod}_{i,j,s\in Y_i} 
\cosh^2{\frac{2\pi b a_{ij}+i\pi b^2 (A_{Y_i}(s) -L_{Y_j}(s)  )}{2} }   ,\NN\\
&& Z_Y^{\rm fund} (a)
 = \hat{\prod}_{j, s\in Y_j} 
 \cosh{\frac{2\pi ba_j +i\pi b^2 ( s_v +s_h -1)}{2}  }   .
 \label{eq:monoContributions}
\end{\eqa}
The sum in \eqref{eq:mono} is over a set of Young tableau 
with the total number of boxes $\frac{1}{2}{\rm tr}(B^2 -v^2)$.
Note also that we have put the symbol $\hat{}\ $ for the sum and product
which means 
the sum over $Y$ and product over $s$ are constrained.
Namely, we include  only $Y$ and $s$ satisfying the constraints
in contrast to the instanton corrections. 
We explain details on this in app.~\ref{app:bubbling}
because the constraints are quite complicated
and nevertheless their details are not so important for our purpose.

The most important thing for us 
is singularity structure of $Z_{\rm mono}^{(v)}(a)$.
For the contribution from single set of Young diagram $Y$, 
the vector contribution gives poles at
\begin{\eq}
a_{ij}
= ib^{-1}m -\frac{ib}{2} \left( A_{Y_i} (s) -L_{Y_j}(s) \pm 1  \right) .
\end{\eq}
When $A_{Y_i}(s) -L_{Y_j}(s) \pm 1 \neq 0$ or $m\neq 0$,
the poles are not located along real axis unless $b^2$ is purely imaginary.
When $A_{Y_i}(s) -L_{Y_j}(s) \pm 1 = 0$,
we have the poles at $a_{ij} =0$
but this type of poles are canceled from other Young diagrams
as long as we consider well-defined 't Hooft loops.
Thus, $Z_{\rm mono}^{(v)}(a)$ does not have singularities along real axis
and the perturbative series is Borel summable along $\mathbb{R}_+$.
Furthermore, the Borel transformation has singularities only along $\mathbb{R}_-$ for $b\in\mathbb{R}$
including all the non-perturbative corrections.

\subsubsection*{Dyonic loop}
Supersymmetric dyonic loop 
can be computed by putting SUSY Wilson loop on $S_b^1$ \cite{Hama:2012bg}
in the setup of the 't Hooft loop.
If we assume the conjectural expression \eqref{eq:MM} for the 't Hooft loop, 
then the exact result for the dyonic loop should be given by
\begin{\eq}
D=\langle {\rm tr}_R e^{ba} \rangle_{M.M.},
\end{\eq}
where $\langle \cdots \rangle_{M.M.}$ denotes
unnormalized expectation value in the matrix model \eqref{eq:MM}.
Since this is just insertion of the function of $a$ without singularities,
this does not prevent us from application of the above technique
to this case.
Then the Borel transformation is simply given 
by $f^{(k,\bar{k},v)}$ in \eqref{eq:f_multi}
with the extra insertion of ${\rm tr}_R e^{ba}$ to the integrand.
Thus, the analytic properties of $h^{(k,\bar{k},v)}$
do not change and
the perturbative series is still Borel summable along $\mathbb{R}_+$
including the instantons and the monopole bubbling effects.

\section{Conclusion and Discussions}
\label{sec:conclusion}
In this paper 
we have studied weak coupling perturbative series
in the presence of monopole bubbling effects as well as instanton effects.
We have shown that
the perturbative series of the 't Hooft loop 
in the 4d $\mathcal{N}=2$ supersymmetric gauge theories
are Borel summable along $\mathbb{R}_+$.
It has turned out that
the exact result is the same as
the sum of the Borel resummations 
over instanton-anti-instanton numbers $(k,\bar{k})$ and screened magnetic charges $v$.
Our result is also non-trivially consistent 
with the conjecture that
4d $\mathcal{N}=2$ theories do not have
IR renormalon type singularities \cite{Argyres:2012vv,Poppitz:2011wy}.

There is a confusing point on our results.
According to Lipatov's argument \cite{Lipatov:1976ny}
and topological selection rule \cite{Argyres:2012vv,Dunne:2012ae} 
(see also \cite{Aniceto:2013fka}),
it is expected that
Borel singularities come from saddle points with the same topological number.
Hence, one may expect 
that there are
Borel singularities coming from instantons-anti-instantons 
with the same $k-\bar{k}$ in our setup.
However, we have seen that we do not have such singularities.
Absence of this type of singularities as well as IR renormalons
lead the Borel summability along $\mathbb{R}_+$ and
makes the perturbative series in every sector isolated
in contrast to usual resurgence scenario 
in quantum mechanics \cite{Bogomolny:1980ur}
and QFT \cite{Argyres:2012vv,Dunne:2012ae,Cherman:2013yfa,Honda:2016vmv}.
It is interesting to find physical interpretations for this point.
One of possible scenarios would be 
``Chesire cat resurgence",
which has recently appeared in some SUSY theories \cite{Kozcaz:2016wvy}.
Another possibility is that
there is something like a generalization of topological selection rule 
particular for our setup.

We have found that
the Borel transformation has infinitely many singularities in Borel plane.
It is interesting to find their physical interpretations.
Technically, these singularities come 
from singularities of the one-loop determinant, Nekrasov partition functions,
and monopole bubbling effects.
At least for those from the one-loop determinant,
we expect that
they can be explained by complexified SUSY solutions
as in 3d $\mathcal{N}=2$ case \cite{Honda:2017qdb,Honda:2016vmv},
which formally satisfy SUSY conditions 
but may not be on the original path integral contour.

The formula for the 't Hooft loop, which we have used,
has not been explicitly computed by the localization
despite it has the passed nontrivial checks.
It is nice to derive the localization formula directly.

It is known that
the 't Hooft loop operator is S-dual to SUSY Wilson loop,
whose perturbative expansion is also Borel summable along $\mathbb{R}_+$ \cite{Honda:2016mvg}.
It would be illuminating to study
whether there are implications of the $S$-duality 
for structures of Borel singularities.

\subsection*{Acknowledgments}
We thank Takuya Okuda for his correspondence
to our questions on \cite{Gomis:2011pf}.
M.~H. would like to thank 
Centro de Ciencias de Benasque,
CERN, Fudan University, KITP, RIKEN, and YITP for hospitality.
D.~Y. would like to thank Tokyo Institute of Technology for hospitality.
This research was supported in part by the National Science Foundation under Grant No. NSF PHY-1125915.
D.~Y. was supported by the ERC Starting Grant N. 304806, ``The Gauge/Gravity Duality and Geometry in String Theory''.

\appendix
\section{Analytic property of one-loop determinant}
\label{app:analytic}
We discuss details on analytic properties of 
the one-loop determinant $Z_{\rm 1loop}^{(v)} (a)$ as a function of $a$.
\subsection*{Hyper multiplet contribution}
We study an analytic property of a contribution from a weight vector $\rho$:
\begin{\eq}
\frac{ \prod_{k=0}^{|\rho \cdot v|-1}
\cosh^{\frac{1}{2}}{\Bigl[ \pi b \left(\rho \cdot a 
+i b \left( k-\frac{|\rho \cdot v|-1}{2}\right) \right) \Bigr]}}
{\left| \Upsilon \left( i \rho \cdot a_N +\frac{Q}{2} \right)  \right| } .
\label{eq:startH}
\end{\eq}
Using the infinite product representation \eqref{eq:Upsilon} of $\Upsilon (x)$,
we have the following convenient identity:
\begin{\eqa}
&&\left| \Upsilon  
\left( i \left( x +\frac{ibv}{2} \right) +\frac{Q}{2} \right) \right|^2 = \NN\\
&&\hspace{-1em}
 \prod_{k =0}^{|v|-1} \prod_{m_2\geq 0 } 
\left[ \left( m_2 +\frac{1}{2}\right)^2 b^{-2}
   +\left( x +ib\left(k -\frac{|v|-1}{2} \right)   \right)^2\right] \NN\\
&&\hspace{-1em}
 \prod_{m_1 ,m_2 \geq 0}
\left[ \left(  \left( m_1 +\frac{|v|+1}{2} \right) b 
 +\left(m_2 +\frac{1}{2} \right) b^{-1} \right)^2 
  +x^2 \right]^2 . \NN\\
\end{\eqa}
Then, recalling
$\cosh{(\pi x)}
=\prod_{n=1}^\infty \left(  1+ \frac{4x^2}{(2n-1)^2}  \right)$,
the first term cancels the equator contribution and 
we find that \eqref{eq:startH} is proportional to
\begin{\eq}
\prod_{m_1 ,m_2 \geq 0}
\frac{1}{ \left(  \left( m_1 +\frac{|\rho\cdot v|+1}{2} \right) b +\left(m_2 +\frac{1}{2} \right) b^{-1} \right)^2   +(\rho \cdot a)^2 } .
\end{\eq}
Thus, we do not have branch cuts but have simple poles at
\begin{\eq}
\rho\cdot a
= \pm i \left[  \left( m_1 +\frac{|\rho\cdot v|+1}{2} \right) b +\left(m_2 +\frac{1}{2} \right) b^{-1} \right]  .
\end{\eq}

\subsection*{Vector multiplet contribution}
Next we study an analytic property of each positive root contribution:
\begin{\eq}
 \frac{\left| \Upsilon (i \alpha\cdot a_N ) \Upsilon (-i \alpha\cdot a_N ) \right| }
{\prod_\pm \prod_{k=0}^{|\alpha \cdot v|-1}
\sinh^{\frac{1}{2}}{\Bigl[ \pi b\left(\pm \alpha \cdot a 
+ib \left( k-\frac{|\alpha \cdot v|}{2} \right) \right)\Bigr]}  } .
\label{eq:startV}
\end{\eq}
For this case, it is convenient to use the identity
\begin{\eqa}
&& \left| 
\Upsilon  \left( i\left( x+\frac{ibv}{2} \right) \right) 
\Upsilon  \left( -i\left( x+\frac{ibv}{2} \right) \right)  \right|^2 \NN\\
&& \propto \prod_\pm \prod_{k =0}^{|v|-1}
\sinh{\Biggl[ \pi b \left( \pm x  +ib\left(k -\frac{|v|-1}{2} \right)   \right)\Biggr] } \NN\\
&&  \prod_{m_1 ,m_2 \geq 0} 
\Biggl[ \left(  \left( m_1 +\frac{|v|+1}{2} \right) b 
 +\left(m_2 +\frac{1}{2} \right) b^{-1} \right)^2 \NN\\ 
&&~~~~~~~~~~~  +\left(\pm x +\frac{iQ}{2}  \right)^2 \Biggr]^2 .
\end{\eqa}
Then the first term cancels the equator contribution and
\eqref{eq:startV} becomes proportional to
\begin{\eqa}
&&\prod_{m_1 ,m_2 \geq 0} 
\Biggl[ \left(  \left( m_1 +\frac{|\alpha \cdot v|+1}{2} \right) b 
 +\left(m_2 +\frac{1}{2} \right) b^{-1} \right)^2 \NN\\
&&~~~~~~~~~  +\left( \alpha\cdot a +\frac{iQ}{2}  \right)^2 \Biggr] ,
\quad {\rm with}\ \alpha \in \Delta .
\end{\eqa}
Thus, the vector one-loop determinant does not have any singularities.
Note also that this has simple zeroes at
\begin{\eq}
\alpha\cdot a
=-\frac{iQ}{2} \pm i \left[  \left( m_1 +\frac{|\alpha \cdot v|+1}{2} \right) b 
 +\left(m_2 +\frac{1}{2} \right) b^{-1} \right] .
\end{\eq}
Thus the whole one-loop determinant does not have branch cuts
and is meromorphic function of $a$
whose poles are not located on real axis.
This point is directly connected to Borel summability along $\mathbb{R}_+$.

\section{Details on monopole bubbling effects}
\label{app:bubbling}
We explain details of the monopole bubbling effect \eqref{eq:mono}.
As we mentioned,
the sum over $Y$ in \eqref{eq:mono} 
and product over $s$ in \eqref{eq:monoContributions}
are constrained.
$Y$ in (\ref{eq:mono}) is a set of Young diagrams ($Y=\{Y_1,\cdots,Y_N\}$),
and the sum is over all possible configurations of $p$ boxes
distributed to $Y$.
The number of the boxes $p$ is determined 
by the dimension of a vector $K=(K_1 ,\cdots ,K_p )$,
which is specified by a following equation \cite{Gang:2012yr}
\begin{align}
 \tr \left( \alpha^B \right) = \tr \left( \alpha^v \right)
 + \alpha^{-1}\left(\alpha-1\right)^2 \tr \left(\alpha^K \right)  ,
\end{align}
where $\alpha$ is arbitrary element of $U(1)$ and
and the trace is the sum over all components of the vector.
Taking $\alpha =e^{i\epsilon}$ and comparing $\mathcal{O}(\epsilon^2 )$,
this condition uniquely determines $p$ as $p=\frac{1}{2}{\rm tr}(B^2 -v^2 ) $.
Especially when $v=B$,
we do not have solutions and
therefore the monopole bubbling effect is trivial 
as mentioned in (\ref{eq:trivialMono}).
The products in terms of $i, j, s$ in (\ref{eq:monoContributions})
have two constraints.
One is
\begin{align}
 v_{i(s)} +s_h -s_v \in \left\{K_t\right\}_{t=1}^p  ,
\end{align}
where $i(s)$ is a gauge group index of Young tableau that $s$ belongs to.
The other constraint depends on the representation.
The constraint for vector and adjoint representations is
\begin{align}
 v_{ij} +L_{Y_j}(s)+A_{Y_i}(s)+1 = 0  ,
\end{align}
while the one for fundamental representation is
\begin{align}
 v_{i(s)} -s_v +s_h = 0  .
\end{align}
In order to find $(B,v)$ for $SU(N)$,
it is convenient to start 
with the vectors $(\tilde{B},\tilde{v})$ 
satisfying the above constraints for $U(N)$
and impose the traceless condition:
\begin{align}
  B = \wt B -\frac{1}{N}\sum_{i=1}^N \wt B_i  ,  \quad
  v = \wt v -\frac{1}{N}\sum_{i=1}^N \wt v_i  .  
\end{align}

\providecommand{\href}[2]{#2}\begingroup\raggedright\endgroup

\end{document}